\documentstyle[pre,aps,eqsecnum,epsf,graphicx]{revtex}
\begin{document}

\draft

\title{Negative specific heat in a thermodynamic model of multifragmentation}
 
\author{C. B. Das$^1$, S. Das Gupta$^1$ and A. Z. Mekjian$^2$}

\address{$^1$Physics Department, McGill University, 
Montr{\'e}al, Canada H3A 2T8}
\address{$^2$Department of Physics and Astronomy, 
Rutgers University, Piscataway, New Jersey 08855}

\date{\today}

\maketitle

\begin{abstract}
We consider a soluble model of multifragmentation which is similar 
in spirit to many models which have been 
used to fit intermediate energy heavy ion collision data.  In this model
$c_v$ is always positive but for finite nuclei $c_p$ can be negative
for some temperatures and pressures.  Furthermore, negative values of $c_p$
can be obtained in canonical treatment.  One does not need to use the
microcanonical ensemble.  Negative values for $c_p$ can persist for systems as
large as 200 paticles but this depends upon parameters used in the model
calculation.  As expected, negative specific heats are absent in the 
thermodynamic limit.

\end{abstract}


\section{Introduction}
This paper deals with specific heats of an assembly of interacting 
nucleons.  In recent times the subject has received a great deal of
attention \cite{Bondorf,Gross,Chomaz,Das,Agostino,Moretto}.  The topic
is beset with many controversies.  Some of the ideas are : 
under suitable condtions, nuclear
systems exhibit negative heat capacities; 
negative heat capacities are obtainable only
in the microcanonical ensemble; negative
heat capacities also appear in canonical models but
disappear once the drop size crosses the value $\approx$ 60.

We investigate the specific heats using a thermodynamic model.  The basic
assumption of the model is that populations of different channels
are dictated solely by phase space considerations.  This is a common
theme in many applications, for example, SMM (statistical
multifragmentation model) \cite{Bondorf} and
MMMC (microcanonical metropolis Monte Carlo)
model \cite{Gross} although details vary from one model to another.

A canonical model based on this assumption was shown to be easily soluble
requiring only very quick and simple computing \cite{Dasgupta1}.  
The first application
used one kind of particle but was later extended to two kinds of particles
\cite{Bhat1,Bhat2}. This appears to be accurate enough for many 
applications \cite{Tsang} and will undoubtedly be used more and more in
the future.  We investigate the question of specific heat in this model
primarily using one kind of particle.  Two kinds of particles were also used
but requires longer computing time but we expect no changes from the lessons
learnt from the model of one kind of particles.  We will however show also
some results obtained from using two kinds of particles.

What we will show is that although $c_v$ for this model is always
positive, $c_p$ can sometimes be negative.  This is a finite particle 
number effect and negative values disappear in the thermodynamic
limit.  Furthermore we get negative values of $c_p$ in the canonical
model itself.  We did not need to go to the microcanonical description.
Thermodynamic limit is obtained by using the grand canonical ensemble
whereas finite systems are described by canonical model with exact
particle number.  We find that negative values of $c_p$ can persist for
fairly large systems although this is dependent upon
binding energies used etc.. This was not investigated in detail.

The statements made above appear to hold for the Lattice Gas Model as 
well.  It was demonstrated that $c_v$ is positve in the Lattice Gas
Model even for a very small system.  This can be shown almost
analytically without having to use a Monte-Carlo simulation \cite{Das}
On the other hand $c_p$ is much harder to calculate in the Lattice Gas
Model.  Chomaz et al find that $c_p$ can be negative in the Lattics Gas
Model \cite{Chomaz}. 

For completeness we describe the canonical thermodynamic model in Section
II.  In the next section we set up the grand canonical model to
get to the thermodynamic limit.  Subsequent sections will show the results.

\section{The Thermodynamic Model}

The thermodynamic model has been described in many places 
\cite{Dasgupta1,Bhat1,Dasgupta2}.  For completeness and to
enumerate the parameters we provide some details.  We describe
the model for one kind of particles only.  The generalisation to two 
kinds could be found in \cite{Bhat1,Dasgupta2}.

If there are $A$ identical particles of only one kind in an enclosure
at temperature $T$, the partition function of the system
can be written as
\begin{eqnarray}
Q_A=\frac{1}{A!}(\omega)^A
\end{eqnarray}
Here $\omega$ is the partition function of one particle.  For a spinless
particle this is $\omega=\frac{V}{h^3}(2\pi mT)^{3/2}$; $m$ is the mass 
of the particle; $V$ is the available volume within which each particle
moves; $A!$ corrects for Gibb's paradox.  If there are many
species, the generalisation is
\begin{eqnarray}
Q_A=\sum\prod_i\frac{(\omega_i)^{n_i}}{n_i!}
\end{eqnarray}
Here $\omega_i$ is the partition function of a composite which has $i$
nucleons.  For a dimer $i=2$, for a trimer $i=3$ etc.  Equation (2.2) is
no longer trivial to calculate.  The trouble is with the sum in the right hand
side of eq. (2.2).  The sum is restrictive.  We need to consider only
those partitions of the number $A$ which satisfy $A=\sum in_i$.  The number
of partitions which satisfies the sum is enormous when A is large.
We can call a
given allowed partition to be a channel.  The probablity of the occurrence
of a given channel $P(\vec n)\equiv P(n_1,n_2,n_3....)$ is
\begin{eqnarray}
P(\vec n)=\frac{1}{Q_A}\prod\frac{(\omega_i)^{n_i}}{n_i!}
\end{eqnarray}
The average number of composites of $i$ nucleons is easily seen from
the above equation to be 
\begin{eqnarray}
<n_i>=\omega_i\frac{Q_{A-i}}{Q_A}
\end{eqnarray}
Since $\sum in_i=A$, one readily arrives at a recursion relation \cite{Chase}
\begin{eqnarray}
Q_A=\frac{1}{A}\sum_{k=1}^{k=A}k\omega_kQ_{A-k}
\end{eqnarray}
For one kind of particle, $Q_A$ above is easily evaluated on a computer for 
$A$ as large as 3000 in matter of seconds.  It is this recursion relation
that makes the computation so easy in the model.  Of course, once one has
the partition function all relevant thermodynamic quantities can be 
computed.

We now need an expression for $\omega_k$ which can mimic the nuclear physics
situation.  We take
\begin{eqnarray}
\omega_k=\frac{V}{h^3}(2\pi mT)^{3/2} k^{3/2} q_k
\end{eqnarray}
where the first part arises from the centre of mass motion of the composite
which has $k$ nucleons and $q_k$ is the internal partition function. For
$k=1$, $q_k=1$ and for $k\ge 2$ it is taken to be
\begin{eqnarray}
q_k=\exp[(W_0k-\sigma(T)k^{2/3}+T^2k/\epsilon_0)/T]
\end{eqnarray}
Here, as in \cite{Bondorf}, $W_0$=16 MeV is the volume energy term, $\sigma(T)$
is a temperature dependent surface tension term and the last term arises
from summing over excited states in the Fermi-gas model.  The value of
$\epsilon_0$ is taken to be 16 MeV.  The explicit expression for $\sigma(T)$
used here is $\sigma(T)=\sigma_0[(T_c^2-T^2)/(T_c^2+T^2)]^{5/4}$ with
$\sigma_0=$18 MeV and $T_c=18$ MeV.  In the nuclear case one might be
tempted to interpret $V$ of eq.(2.6) as simply the freeze-out volume but
it is clearly less than that; $V$ is the volume available to the particles
for the centre of mass motion.  Assume that the only interaction between
clusters is that they can not overlap one another.  This assumption
restricts the validity of the model to low density limit
as was stressed in all previous applications of of the model. 
In the Van der Waals spirit we take
$V=V_{freeze}-V_{ex}$ where $V_{ex}$ is taken here to be constant and
equal to $V_0=A/\rho_0$.  The precise value of $V_{ex}$ is inconsequential
so long as it is taken to be constant.  Calculations employ $V$; the value
$V_{ex}$ enters only if results are plotted against $\rho/\rho_0=
V_0/(V+V_{ex})$ where $\rho$ is the freeze-out density.

In the past, calculations with one kind of particle used the 
parametrisation of eq. (2.7) for all $k$'s however
large.  This means that if the system has $A$ nucleons, the largest
possible cluster allowed in the system also has $A$ nucleons.  While
we will show a few cases with this specification we will also 
consider a variation.  We will
take the value of $q_k$ to be given by eq. (2.7) up to a limit $k=N$
and zero afterwards.  When $A>N$, this simply means that the largest
cluster has $N$ nucleons.

Using standard definitions: 
$E=T^2\frac{\partial lnQ_A}{\partial T}$ and
pressure $p=T\frac{\partial lnQ_A}{\partial V}$ we arrive at
\begin{eqnarray}
E=\sum <n_k>(E_k^{kin}+E_k^{int})
\end{eqnarray}
where $E_k^{kin}=\frac{3}{2}T$ and $E_k^{int}=k(-W_0+T^2/\epsilon_0)+
\sigma(T)k^{2/3}-T[\partial \sigma(T)/\partial T]k^{2/3}$.  The last
term in $E_k^{int}$ was neglected in \cite{Dasgupta1}.  It is included
here but makes little difference.  The expression for pressure is
\begin{eqnarray}
p=\frac{T}{V}\sum<n_i>
\end{eqnarray}
Multiplicity $m$ is given by $m=\sum <n_i>$.

\section{Infinite matter limit: grand canonical model}

If there were only monomers in the grand canonical ensemble
one would solve
\begin{eqnarray}
\rho=\exp(\mu/T)\tilde{\omega}_1
\end{eqnarray}
where $\tilde{\omega}_i=\omega_i/V$ of section II.
Given $\rho$ and $T$ one then finds the chemical potential $\mu$.  The number
of particles is then given by $A=\rho V$ where $V$ and $A$ are very large
(thermodynamic limit).  The fluctuations in the number of particles
implied by the use of grand canonical ensemble are then negligible compared
to the average number $A$.

If we have a model where the only allowed species are monomers and dimers and
the total particle number is very large one would solve:
\begin{eqnarray}
\rho=\exp(\mu/T)\tilde{\omega}_1+2\times\exp(2\mu/T)\tilde{\omega}_2
\end{eqnarray}
where phase-space consideration has implied that chemical equilibration
exists, that is, the chemical potential of the dimer is twice
that of the monomer, i.e., $\mu_2=2\mu$.

For a system which is very large but, for which, the heaviest cluster has
$N$ nucleons and no more, one needs to solve
\begin{eqnarray}
\rho=\sum_1^Nk\exp(k\mu/T)\tilde{\omega}_k
\end{eqnarray}
In this case, one might argue that one is 
considering a model in which the composites obey eq.(2.6) up to $k=N$
and $\omega_k$'s for $k>N$ are all zeroes.  Of course it is possible
that both $A$ and $N$ are very large.  Use of the grand canonical ensemble
always implies that $A$ is very large but $N$ may be large or small.

Pressure in the grand canonical
model is calculated from $p=(T/V)lnZ_{grand}$ which in this model reduces
to $p=(T/V)\sum <n_i>$ where $<n_i>/V=\exp(i\mu/T)\tilde{\omega}_i$.  Notice
that formally this eq. is same as eq.(2.9) but, of course, $<n_i>$ in eq.
(2.9) is calculated according to the canonical formula, eq.(2.4).

\section{specific heats in the model}

In \cite{Dasgupta1} where the canonical thermodynamic model was first studied
for phase transitions, it was pointed out that for a given density
$\rho$, the specific heat per particle $c_v=C_V/A$ tends to $\infty$ at a 
particular temperature when the particle number $A$ tends to $\infty$.
Since,
\begin{center}
\begin{math}
C_V=\left(\frac{\partial E}{\partial T}\right)_V = T \left(\frac{\partial S}{\partial T}\right)_V
 = -T \left(\frac{\partial^2F}{\partial^2T}\right)_V, \nonumber
\end{math}
\end{center}
a singularity in $C_V$ signifies
a break in the first derivative of $F$, the free energy and a first
order phase transition.  The specific heat $c_v$ in the model has been
studied in more than one application and always found to be positive.  We now 
turn our attention to $c_p$ studied in this canonical model.  It is instructive
to look at $p-\rho$ curves at different temperatures (equation of state) to 
gain an understanding (Fig.1).
For 200 particles ($A$=200 and $N$=number of 
nucleons of the largest allowed cluster=200)
this is drawn at three temperatures: $T_1<T_2<T_3$.  Here
$T_2$ is only slightly higher than $T_1$.  We notice that on $T_1, T_2$
isothermals there are regions of mechanical instability where $\partial p
/\partial \rho$ is negative.  It is in this region that one encounters negative
values for $c_p$.  Instead of $\rho$ let us use the variable 
$V\propto 1/\rho$.  Thus regions of mechanical instability are characterised by
$\partial p/\partial V>$0.  Let us try to understand how this can happen.
In simpler cases as in a gas of noninteracting monomers, the multiplicity
$m$ (which determines the pressure, i.e., $p=T\times\frac{m}{V})$ is simply 
$A$ and 
$(\partial p/\partial V)_T$ is always negative.  
(we actually use $m-1$ rather than $m$ for calculating $p$ but this
is immaterial for our discussion).   In the thermodynamic model, because
of composites, $m<<A$ at moderate temperatures.
At fixed temperature $T>0$,  $m$ will always increase
with $V$. Negative compressibility is marked by 
$(\frac{\partial m}{\partial V})_T> \frac{m}{V}$.  Let us consider points
$a,b$ (region of positive $c_p$) and points $c,d$ (region of negative $c_p$).
Points $b,c$ are on $T_1\equiv T$ and points $a,d$ are on 
$T_2\equiv T+\delta T$ with $\delta T>0$.  Using

$p=T\frac{m}{V}=(T+\delta T)\frac{m+\delta m}{V+\delta V}$ we arrive at
\begin{eqnarray}
\frac{\delta m}{m}=\frac{\delta V}{V}-\frac{\delta T}{T}
\end{eqnarray}
In the region $(c,d)$, $\delta V$ is negative, $\delta T$ ia always positive
thus $\delta m$ is negative.  If $m$ goes down then so does the 
kinetic energy
and also the potential energy (creating more $m$ creates more surface
and hence more energy).  Thus in this region with increasing temperature
but constant pressure,
both the kinetic and the potential energy of the system go down.  In the
``normal'' region $\delta m$ is positive and both the kinetic and 
the potential energies increase with $T$ at constant pressure.  This is illustrared
in Table 1.  Finally we show, in fig. 2, the caloric curve for a given 
pressure where
in part of the curve temperature does go down with excitation energy.
The fall is very gentle whereas the rise with energy when it happens is
faster.

The occurrence of a negative $C_p$ in spite of a positive $C_V$ is
allowed in the following well-known relation \cite{Reif}:
\begin{center}
\begin{math}
C_p-C_V=VT \frac{\alpha^2}{\kappa} 
\end{math}
\end{center}
where $\alpha$ is the volume
coefficient expansion and $\kappa$ is the isothermal compressibility
given by:
\begin{eqnarray}
\alpha &=&\frac{1}{V} \left(\frac{\partial V}{\partial
T}\right)_p \nonumber \\
\kappa &=& -\frac{1}{V} \left(\frac{\partial V}{\partial p}\right)_T. \nonumber
\end{eqnarray}
For
negative
$\kappa$, $C_p$ is less than $C_V$ and can become negative.

Using the equality,
\begin{center}
\begin{math}
\left(\frac{\partial V}{\partial T}\right)_p = -\left(\frac{\partial V}{\partial p}\right)_T \left(\frac{\partial p}{\partial T}\right)_p
\end{math}
\end{center}
we can also write,
\begin{center}
\begin{math}
C_p-C_V=VT \left(\frac{\partial p}{\partial
T}\right)_V\frac{1}{V}\left(\frac{\partial V}
{\partial T}\right)_p .
\end{math}
\end{center}
This shows that, $C_p$ can drop below $C_V$ if 
isobaric volume coefficient of expansion becomes negative which is
the case in some regions of Fig.1.

\section{Extrapolation to thermodynamic limit}
For $A$ very large but N=200 we use the grand canonical ensemble.
For a given $\rho$ and $T$, we solve for $\mu$ where $N$ in eq. (3.3)
is set at 200.  In the $p-\rho$ diagram there are no regions of mechanical
instability (see fig. 3).  For comparison, the $p-\rho$ diagram for
$N=200$ but $A$=200 obtained by the canonical calculation is also
shown in the same figure.  We see that in the low density side (the
gas phase) the two diagrams coincide.  The rise of pressure with density
is quite rapid and linear.  After the two diagrams separate, the rise of
pressure with density in the grand canonical model slows down considerably
but there is no region of mechanical instability although the canonical
calculation with 200 particles has a region of instability.  In the
grand canonical result which represents the thermodynamic extrapolation,
we have not reached the classic liquid-gas coexistence limit where
there would be no rise of pressure at all (like in Maxwell construction).
We think the reason is this.  The largest cluster has size 200 
which is not a big enough number.  Condensation
into the largest and larger clusters still does not behave like a liquid.
We now increase the largest cluster size to 2000.  Now the coexistence
region is very clear and there is unmistakable signature of first order
phase transition.  This grand canonical result is very close to the
case where $N=\infty$ ( see Fig. 3 in \cite{Bugaev}).
In the same figure we also show results of a canonical
calculation with $A$=2000 and $N$=2000.  The region of mechanical 
instability has gone down considerably but it has not disappeared showing
that we have not reached the thermodynamic limit yet.

\section{More calculations in the canonical model}
The mechanical instability which led to negative values of $c_p$ is not
only a finite number effect but it is also dependent on details of
parameters; see also \cite{Moretto}.  As in fig.3 we draw a $p-\rho$
diagram for 200 particles in fig. 4 but now the largest cluster has 
$N=100$, that is, $\omega_k$ is given by eq. (2.6) up to $k$=100 and is 
zero for $k>100$.  The mechanical instability region has completely 
disappeared.  In fact, the negative compressility in the $p-\rho$ diagram 
in fig. 4 disappears even with the following minimal change.  We use $q_k$ of
eq. (2.7) upto $k=100$ and for $k>100$ use 
$q_k=\exp[0.97(W_0k-\sigma(T)k^{2/3}+T^2k/\epsilon_0)/T]$. We
were surprised that with such small changes zones of negative
compressibility disappeared but such consequences were anticipated
in other models before \cite{Moretto}.

\section{Chemical Potentials}
In this section we deal with two kinds of particles and discuss the 
behaviours of chemical potentials as the proton fraction of large
systems changes.  This is remotely connected with specific heats
but the behaviour of chemical potentials with the proton fraction
has attracted some attention in recent times and we felt that it is
of general interest to show what the behaviour is in the
themodynamic model.  It was shown that
in mean-field theories of nuclear matter there is chemical
instability
as a function of $y=Z/(N+Z)$ in limited regions of $y$, that is,
$({\partial \mu_P}/{\partial y})_{p,T}$
becomes negative in some region
of $\mu_P-y$ plane (correspondingly 
$({\partial \mu_N}/{\partial y})_{p,T}$
becomes positive).
This is analgous to mechanical instability as a function of density
\cite{Muller1,Muller2,Lee}.  We have seen that in the thermodynamic
model there are no regions of mechanical instability for large
systems (the grand canonical results).  We will see that there is 
no chemical instability either in the model in the large particle
number
limit.
Now we need to consider the thermodynamic model for two kinds of
particles.
For details we refer to \cite{Bhat1}. A composite has two
indices:
$i$=proton number, $j$=neutron number with $a=i+j$.  Analogous to
eq.(2.4)
we have
\begin{eqnarray}
<n_{i,j}>=\omega_{i,j}\frac{Q_{Z-i,N-j}}{Q_{Z,N}}
\end{eqnarray}
where the nuclear properties are contained in $\omega_{i,j}$:
\begin{eqnarray}
\omega_{i,j}=\frac{V}{h^3}(2\pi mT)^{3/2} a^{3/2} q_{i,j}
\end{eqnarray}
We take the internal partition function of the composite to be
\begin{eqnarray}
q_{i,j}=\exp[(Wa-\sigma
a^{2/3}-s\frac{(i-j)^2}{a}+aT^2/\epsilon_0)/T]
\end{eqnarray}
As is usual in all infinite matter case calculations,
the coulomb interaction is switched off.
We take $W=15.8$ MeV, $\sigma=18$ MeV,$s$=23.5 MeV and
$\epsilon_0$=16.0
MeV.  For $a\ge 5$ we use this formula.  For lower masses we simulate
the
no coulomb case by setting the binding energy of $^3$He=binding
energy of
$^3$H and binding energy of $^4$Li=binding energy of $^4$H.

For a given $a$ what are the limits on $i$(or $j=a-i$)?  This is a 
non-trivial question.  In the results we will show, we have taken the
limits by calculating the drip lines of protons and neutrons as given
by the binding energy formula.  Limiting oneself to within the drip
lines
is a well-defined prescription, but is likely to be an
underestimation
since resonances show up in particle-particle correlation
experiments.
On the other hand, for a given $a$, taking limits of $i$ from 0 to
$a$ is definitely an overestimation.

In fig. 5 we have drawn isothermals (at $T$=6.0 MeV)
for two component nuclear systems for different $y$'s
in the grand canonical ensemble.  We restrict $y$ between 0.3
and 0.5 and $\rho/\rho_0$ between 0 and 0.5, the ranges for which 
the thermodynamic model is expected to be reliable.  In the
calculation
the largest cluster is taken to be $a=500$.  The same figure also
shows the behaviours of $\mu_P$ and $\mu_N$ at constant pressure.
The derivative $\frac{\partial \mu_p}{\partial y}$ is seen to be 
always positive (simultaneously $\frac{\partial \mu_N}{\partial y}$
is negative).  In the next figure, for completeness, we have
continued 
the model beyond $\rho/\rho_0$=0.5 and gone upto the highest possible
limit of $\rho/\rho_0=1$ in the model to see the behaviour of
$\mu_P$ and $\mu_N$.  No chemical instability is seen.

\section{Summary and Discussion}

We have shown that with usual concepts one can obtain a negative
value of $C_P$ in part of the $T-E$ plane within the framework
of a thermodynamic model.  Although we have shown this, for the sake
of simplicity, using one kind of particle only, we have checked that
the phenomenon remains when a more complicated version with two
kinds of particles and realistic binding energies for the
composites are used.  The $C_V$ is positive and its origin is
the cost in surface energy to break large clusters into smaller
clusters and nucleons.  A negative $C_p$ is seen in our exactly
soluble canonical ensemble model for small systems.
This negative value arises in regions of mechanical instability
where the isothermal compressiblity is negative or equivalently,
the isobaric volume expansion coefficient is negative.  A negative
isobaric volume expansion leads to a decrease in multiplicity,
or total number of clusters, with temperature and a corresponding
decrease in energy.  For larger systems these regions disappear
and in the grand canonical limit, $C_p$ is always positive.

Since several papers have demonstrated the
existence of negative specific heats it is pertinent to mention
the relevance of our work to these earlier works.  Our model
is not, in any simple way, connected to negative specific heat
found in ten dimensional Potts Model \cite{Gross}.  The specific 
heat considered in that work is $C_V$ and the negative specific
heat appears only in microcanonical treatment.  The negative specific 
heat seen here is at least partially similar to that seen in \cite 
{Chomaz,Das}.  There is no negative specific heat in $C_V$ \cite{Das}
but it makes its presence felt when $C_P$ is considered
\cite{Chomaz}.
Although our model is quite different from the one considered in 
\cite{Moretto} the results are similar.

Unfortunately we can not recommend any experiments to verify the
conclusions of this paper.  Nuclear disasembly in heavy ion
collisions
can not be fine tuned.  There is no reason to think that it takes
place
exactly at constant volume or exactly at constant pressure. 
Calculations
at constant volume gives quite reasonable predictions for observables
that have been measured \cite{Tsang,Majumder}
but this does not rule out the possibility
that variations happen.  If disassembly always took place at constant
pressure then the following idealised experiment would be useful.
One measures excitation energy per particle and also the temperature.
One would then find there are cases where the average excitation 
energy per particle goes down even though the temperature rises.

\section{Acknowledgment}
S. Das Gupta acknowledges very useful discussions with Rajat K. Bhaduri,
Lee Sobotka and Abhijit Majumder.
This work is supported in part by the Natural Sciences and Engineering
Research Council of Canada and the U.S. Department of Energy 
Grant No. DE FG02-96ER40987.

\begin{table}
\caption{Variation of energies per particle $(MeV)$ with 
temperature $(MeV)$ in the negative and positive compressibility 
zones, for $p = 0.017 \ MeV \ fm^{-3}$}
\vspace {0.5in}
\begin{tabular}{cccccc}
\hline
\multicolumn{1}{c}{} &
\multicolumn{1}{c}{$T$} &
\multicolumn{1}{c}{$\rho/\rho_0$} &
\multicolumn{1}{c}{$e_k/A$} &
\multicolumn{1}{c}{$e_{pot}/A$} &
\multicolumn{1}{c}{$e_{tot}/A$} \\
\hline
&6.0&0.146&0.978&-5.235&-4.257 \\
$\frac{\partial p}{\partial \rho}  < 0$&6.1&0.212&0.638&-6.970&-6.332 \\
&6.2&0.392&0.294&-8.708&-8.414 \\
\hline
&6.0&0.104&1.422&-3.271&-1.849 \\
$\frac{\partial p}{\partial \rho}  > 0$&6.1&0.090&1.653&-2.513&-0.859 \\
&6.2&0.082&1.824&-2.027&-0.202 \\
\hline
\end{tabular}
\end{table}

\begin{figure}
\epsfxsize=5.5in
\epsfysize=7.0in
\centerline{\epsffile{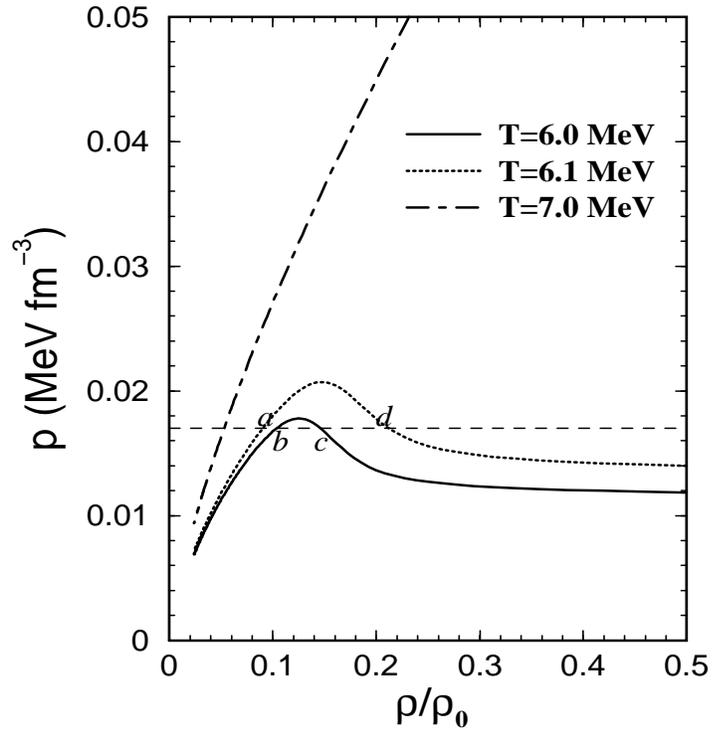}}
\caption{EOS in the canonical model for a system of $A$=200. The
largest cluster also has $N$=200.}
\end{figure}

\begin{figure}
\epsfxsize=5.5in
\epsfysize=7.0in
\centerline{\epsffile{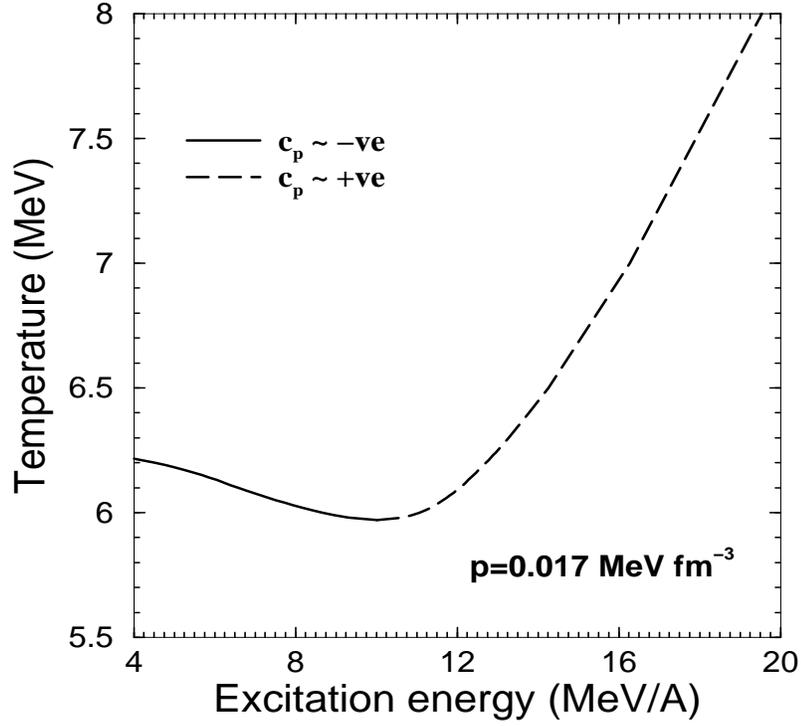}}
\caption{Caloric curve at a constant pressure 
$(p = 0.017 \ MeV \ fm^{-3})$ in the canonical model with $A$=200
and $N$=200. The solid and dashed portions of
the curve give -ve and +ve $c_p$ respectively.}
\end{figure}

\begin{figure}
\epsfxsize=5.5in
\epsfysize=7.0in
\centerline{\epsffile{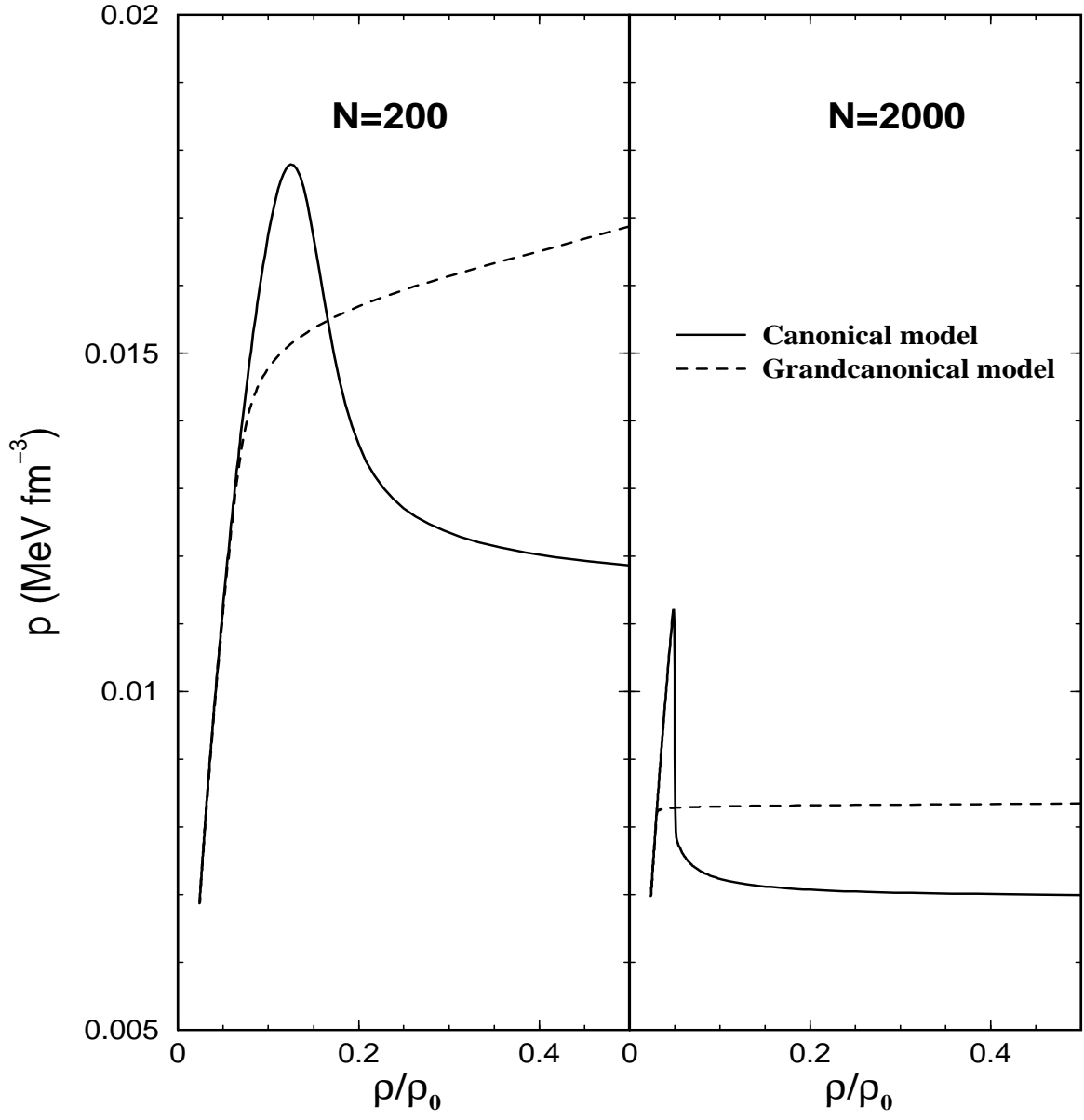}}
\vskip 0.2in
\caption{EOS at $T = 6 \ MeV$ in the two models. For the 
left panel the largest cluster has $N$=200 and for the right
panel $N$=2000. For the canonical
calculation, the left and right panel has $A$=200 and 2000 
respectively, but for the grandcaonocal calculations, $A = \infty$.
(See text).}
\end{figure}

\begin{figure}
\epsfxsize=5.5in
\epsfysize=7.0in
\centerline{\epsffile{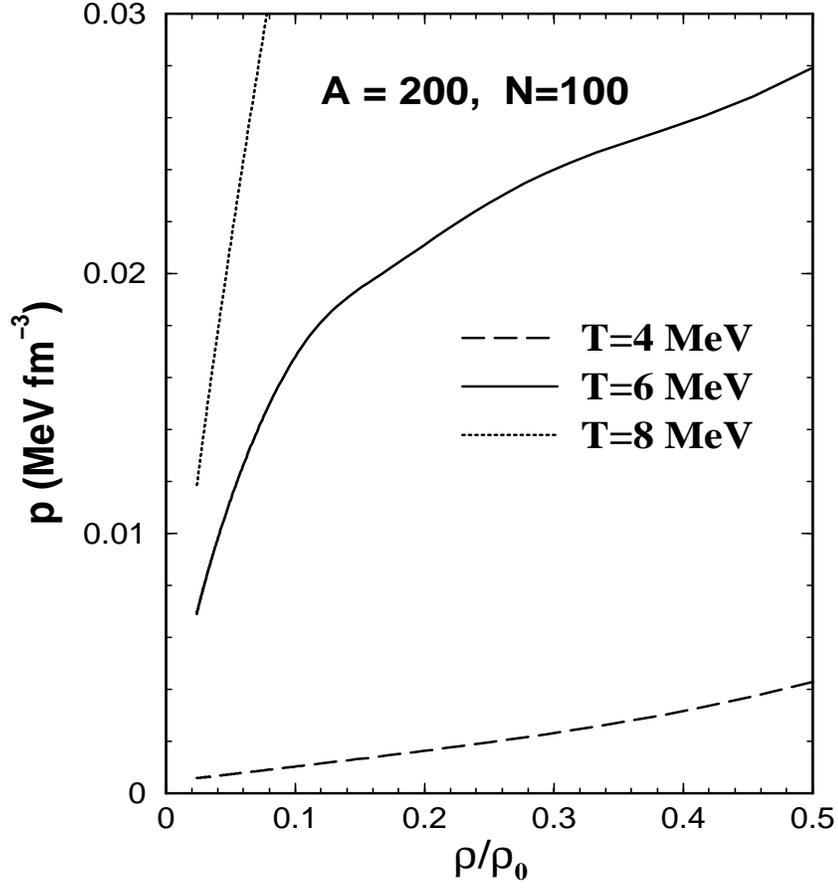}}
\caption{EOS in the canonical model for a system of 200 particles,
but the number of nucleons of the largest cluster is restricted to
100.}
\end{figure}

\begin{figure}
\epsfxsize=5.5in
\epsfysize=6.8in
\centerline{\epsffile{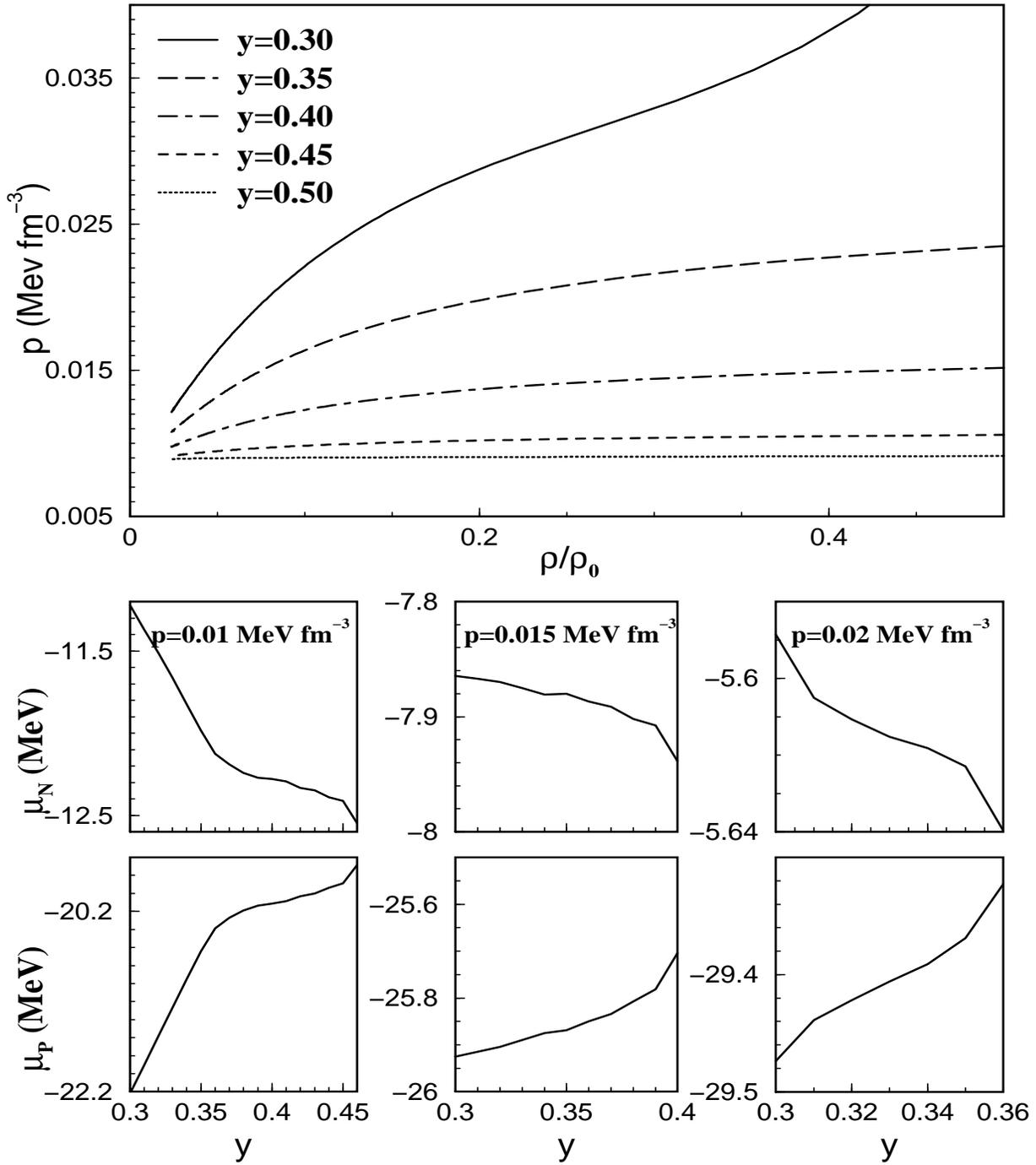}}
\vskip 0.8in
\caption{The top panel are isothermals at $T = 6 MeV$ for different
$y$'s (proton fraction). The lower panels show the behavior of
$\mu_N$ and $\mu_P$ as a function of $y$ in the density range
of the top panel. Calculations are done in a grandcanonical
model with the largest cluster having 500 nucleons.}
\end{figure}

\begin{figure}
\epsfxsize=5.5in
\epsfysize=7.0in
\centerline{\epsffile{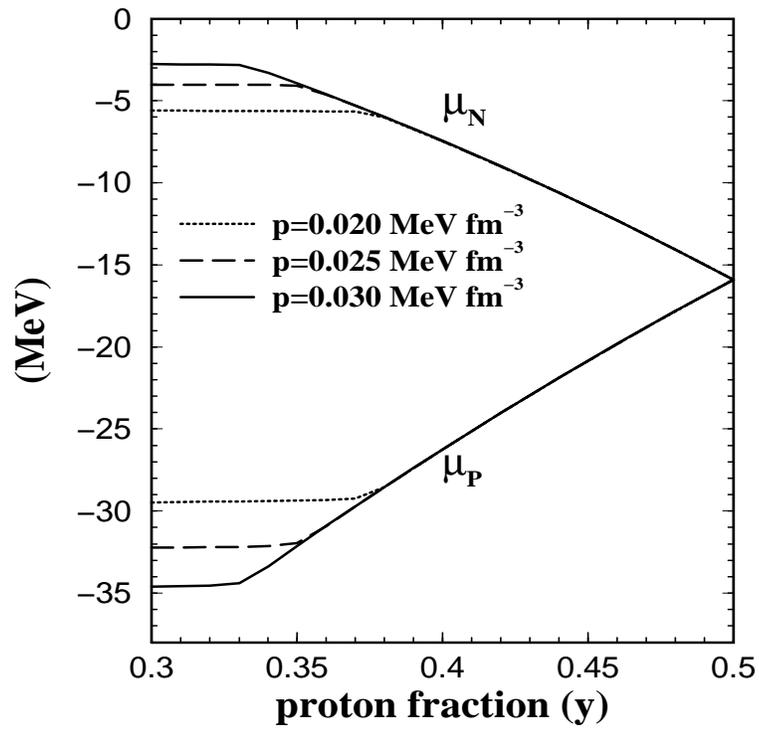}}
\caption{The behaviors of $\mu_N$ and $\mu_P$ extrapolated to
higher densities. For this figure the $p-{\rho/\rho_0}$ diagram (upper
panel Fig.5) was extended upto $\rho/\rho_0$=1.}
\end{figure}

\end{document}